# The model of information retrieval based on the theory of hypercomplex numerical systems

D.V. Lande, Ya.A. Kalinovskiy, Yu.E. Boyarinova

Department of specialized modelling tools
Institute for information recording, Kiev, Ukraine

**Abstract**

The paper provided a description of a new model of information retrieval, which is an extension of vector-space model and is based on the principles of the theory of hypercomplex numerical systems. The model allows to some extent realize the idea of fuzzy search and allows you to apply in practice the model of information retrieval practical developments in the field of hypercomplex numerical systems.

**Keywords:** hypercomplex numbers systems, information retrieval, logical operations, vector-space model, table of multiplication of hypercomplex numbers, estimation of proximity, information presenting.

Currently there are several traditional models of information retrieval: boolean, probabilistic, vector-space [1]. Here's another model based on the use of hypercomplex numerical systems (HNS) and is a development of the canonical vector-space model. The above approach allows to consider the logical objection operation, the phenomenon of synonymy, homonymy etc.

Today HNS are used in solving problems in various fields: theoretical physics [2,3], navigation, guidance and control of movement in three-dimensional space [4,5], computer graphics [6], searching and filtering multimedia [7,8]. Dual numbers, dual quaternions have been used in problems of modelling and control robots and manipulators with many degrees of freedom, in the problems of biomechanics etc. [9-11].

Go to the formal presentation. Let the vocabulary words and phrases constants (terms) contains $N$ terms: $T = \{t_1, t_2, ..., t_N\}$. According to the classical vector-space model, document represented as a vector of weighted terms of values and dimension $N$: $\vec{d} = (w_1, w_2, ..., w_N)$. This weight value can be either the number 0 or 1 (term is included or not included in the document) and the weight values in the interval [0,1], which calculated, such as $tf * idf$.

In the proposed model of information retrieval, based on HNS, each of terms assigned two values positive $w_i^+$ and negative $w_i^-$ that meet the entry weight the terms ($w_i^+$) or failure ($w_i^-$). Both values can take significance in he interval [0,1]. In some cases, may suggest: $w_i^+ + w_i^- = 1$.

Consider the possibility of HNS for information retrieval. Suggested use HNS dimension of the $2N$ with basis $\{e_1, e_2, ..., e_{2N}\}$ and the law of multiplication as the next table:

|  | $e_1$ | $e_2$ | $e_3$ | $e_4$ | ... | $e_{2N-1}$ | $e_{2N}$ |
|---|---|---|---|---|---|---|---|
| $e_1$ | $e_1$ | $e_2$ | 0 | 0 | ... | 0 | 0 |
| $e_2$ | $e_2$ | $e_1$ | 0 | 0 | ... | 0 | 0 |
| $e_3$ | 0 | 0 | $e_3$ | $e_4$ | ... | 0 | 0 |
| $e_4$ | 0 | 0 | $e_4$ | $e_3$ | ... | 0 | 0 |
| ... | ... | ... | ... | ... | ... | ... | ... |
| $e_{2N-1}$ | 0 | 0 | 0 | 0 | 0 | $e_{2N-1}$ | $e_{2N}$ |
| $e_{2N}$ | 0 | 0 | 0 | 0 | 0 | $e_{2N}$ | $e_{2N-1}$ |

Thus, the model of the document is regarded as hypercomplex number that looks like:

$$d = e_1 w_1^+ + e_2 w_1^- + e_3 w_2^+ + e_4 w_2^- + ... + e_{2N-1} w_N^+ + e_{2N} w_N^-.$$

This hypercomplex system is canonical by Shtudi [12] with the unit element $E = e_1 + e_3 + ... + e_{2N-1}$.

In terms of general algebra the hypercomplex number system is a ring that is isomorphically embedded in a full ring of matrix. HNS is commutative and associative $e_i \cdot e_j = e_j \cdot e_i$, $\forall\ i, j = 1, ..., 2N$ and $(e_i \cdot e_j) \cdot e_k = e_i \cdot (e_j \cdot e_k)$, $\forall\ i, j, k = 1, ..., 2N$ [13].

By analogy with the scalar product in Euclidean vector space is considered such an assessment proximity between documents (corresponding hypercomplex numbers) $A = e_1 a_1^+ + e_2 a_1^- + ... + e_{2N-1} a_N^+ + e_{2N} a_N^-$ and $B = e_1 b_1^+ + e_2 b_1^- + ... + e_{2N-1} b_N^+ + e_{2N} b_N^-$:

$$Sim(A, B) = Est\left(\sum_{i=1}^{N} \left(e_{2i-1} a_i^+ + e_{2i} a_i^-\right)\left(e_{2i-1} b_i^+ + e_{2i} b_i^-\right)\right),$$

where $Est()$ – is some additive function estimation hypercomplex numbers. $Est(e_{2i-1}) = e_1$; $Est(e_{2i}) = -e_1$.

According to the proposed model request $Q$ is submitted within the same form as the document $D$. The higher the value of $Sim(Q, D)$, the document more relevant query. It should be noted that the value can be both positive and negative (if we put as usual $e_1 = 1$).

You can type and another assessment of the proximity between hypercomplex numbers, similar to the normal differences between vectors in a vector space:

$$Sim_1(A,B) = Est\left(\sum_{i=1}^{N}\left(e_{2i-1}a_i^+ - e_{2i-1}b_i^+\right)^2 \left(e_{2i}a_i^- - e_{2i}b_i^-\right)^2\right)$$

In this case the document is more relevant to the query if the score $Sim_1(Q,D)$ is lower. At the same time, the second rating the content is less suitable for the tasks of information retrieval. For example, when comparing the document with the query, even with all zero values of the coefficients in the amount requested in the above expression for $Sim_1(Q,D)$ is not zero that is totally dependent on the document.

Therefore, confine the use of the first evaluation. Consider some special cases:

Let the query looks like: $Q = 1/2\,e_1 + 1/2\,e_2$, that is requested word that can be a document or to enter with equal probability. Let the word is a document, namely: $D = e_1$. In this case, $Sim(Q,D) = 1/2 - 1/2 = 0$ which corresponds to complete uncertainty.

Let the query looks like: $Q = 1/2\,e_1 + 1/2\,e_2$, that is requested word that can be a document or to enter with equal probability. Let the paper also looks like $D = 1/2\,e_1 + 1/2\,e_2$. In this case, that $Sim(Q,D) = 1/4 + 1/4 - 1/2 - 1/2 = 0$, as in a previous case corresponds to complete uncertainty.

Let the query has the form $Q = e_1$ and the document: $D = 4/5\,e_1 + 1/5\,e_2$, then $Sim(Q,D) = 4/5 - 1/5 = 3/5$.

The value of the coefficients of the basic elements of the image of a document may match probabilities entry (or entry) terms in a document that can be determined by the probability of errors when writing certain terms, synonyms, homonyms or possible existence.

Application of information retrieval model based on the use of HNS should provide:

- Implementation of logical operations objection that extends canonical vector-space model.
- Implementation of capabilities by fuzzy search, for example, considering the probability of error, as well as training systems, taking into account synonymy, homonymy at expanding multiplication tables corresponding HNS or expand the original query.
- Ability to use existing developments in the field of hypercomplex numerical systems, for example application of finding isomorphic HNS, invariant primary, that more suitable for calculations.

**References**

[1] *Salton G., Wong A., Yang C.S.* A Vector Space Model for Automatic Indexing // Communications of the ACM, 1975. – 18. – № 11. – P. 613–620.
[2] *Mikhailichenko G.G.* Hypercomplex numbers in the theory of physical structures / G.G. Mikhailichenko, R.M. Muradov // Russian Mathematics (Iz VUZ). – 2008. – **52**. – № 10. – P. 20-24

[3] *Paul C.* An octonion model for physics / Paul C. // Proceed. Of ECHO IV Conf. – Odense (Denmark). – 2000.

[4] *Kim M., Hsieh C., Wang M., Wang C., Fang Y., Woo T.* Noise Smoothing for VR Equipment in the Quaternion Space // Proceedings of the Virtual Reality in Manufacturing Research and Education, Department of Industrial Engineering, University of Washington, 1996.

[5] *Ignagni M.B.* On the Orientation Vector Differential Equation in Strapdown Inertial Systems / M.B. Ignagni // IEEE Transactions on Aerospace and Electronic Systems, 1994. – **30**. – № 4. – P. 1076-1081.

[6] *Mukundan R.* Quaternions: From Classical Mechanics to Computer Graphics, and Beyond / R. Mukundan // Proceedings of the 7th Asian Technology Conference in Mathematics, 2002. – P.97-106.

[7] *Villegas M., Paredes R.* Face Recognition in Color Using Complex and Hypercomplex Representations // Proceeding IbPRIA '07 Proceedings of the 3rd Iberian conference on Pattern Recognition and Image Analysis, Part I. – Springer-Verlag Berlin, Heidelberg, 2007.

[8] *Al-Qadda F. S., Ghouti L.* Color Texture Retrieval Using Hypercomplex Wavelets // Proceeding 2009 Symposium on Bio-inspired Learning and Intelligent Systems for Security, BLISS 2009, Edingburgh, United Kingdom, August 20-21 2009, IEEE Computer Society, 2009. – P. 121-126.

[9] *Sangwine S. J.* Fundamental representations and algebraic properties of biquaternions or complexified quaternions / S. J. Sangwine, Todd A. Ell, Nicolas Le Bihan, 2010. Available at http://arxiv.org/abs/1001.0240v1.

[10] *McCarthy J.* Dimensional Synthesis Robots using a Double Quaternion Formulation of the Workspace / J. McCarthy , S. Ahlers // Robotics Research: The Ninth International Symposium, 2000. – P. 3-8.

[11] *Yefremov A.P.* Quaternions: algebra geometry and physical theories / Yefremov A.P.// Hypercomplex numbers in geometry and physics, 2004. – **1**. – P. 104-120

[12] *Study E.* Űber Systeme von complexen Zahlen // Nachrichten von der K.G.D.M. zu Gottingen, 1889. – № 9. – P.237-268.

[13] *Sinkov M.V.* Finite-dimensional hypercomplex number systems. Fundamentals of the theory. Applications. / M.V. Sinkov, Yu.E. Boyarinova, Ya.A. Kalinovskiy. – Kiev: Infodruk, 2010. – 388 p.